\documentclass[9pt,twocolumn,twoside]{opticajnl}
\journal{opticajournal} 
\setboolean{shortarticle}{true}

\newcommand{\hb}{\mathbf{h}}
\newcommand{\rb}{\mathbf{r}}
\newcommand{\Jb}{\mathbf{J}}
\newcommand{\pb}{\mathbf{p}}
\newcommand{\qb}{\mathbf{q}}

\title{Extending the capabilities of vectorial ptychography to circular-polarizing materials such as cholesteric liquid crystals}

\author[1,*]{Patrick Ferrand}
\author[2]{Michel Mitov}

\affil[1]{Aix-Marseille Univ, CNRS, Centrale Méditerranée, Institut Fresnel, F-13013 Marseille, France}
\affil[2]{Univ. Côte d'Azur, CNRS, Institut de Physique de Nice, INPHYNI,  F-06200 Nice, France}
\affil[*]{patrick.ferrand@fresnel.fr}

\begin{abstract}
The problem of imaging materials with circular polarization properties is discussed within the framework of vectorial ptychography. We demonstrate, both theoretically and numerically, that using linear polarizations to investigate such materials compromises the unicity of the solution provided by this computational method. To overcome this limitation, an improved measurement approach is proposed, which involves specific combinations of elliptical polarizations. The effectiveness of this strategy is demonstrated by numerical simulations and experimental measurements on cholesteric liquid crystals films, which possess unique polarization properties. With the help of Pauli matrices algebra, our results highlight the technique's ability to discern between different types of circular polarizers, uniform vs. non-uniform, and determine their handedness.
\end{abstract}

\setboolean{displaycopyright}{false} 

\begin{document}

\maketitle

\section{Introduction}

Vectorial ptychography is a recent imaging technique that uses phase retrieval algorithms to provide quantitative maps of Jones matrices \cite{ferrand_2015}. It is used in optical microscopy to study specimens that strongly affect the phase and polarization of transmitted light, and it has a robust reference-free experimental scheme \cite{ferrand_2018}. This variant of optical ptychography \cite{wang_optical_2023} has been successfully applied to various materials, ranging from natural substances like biomineral calcareous shells \cite{baroni_prapp_2020,duboisset_2022} or biological tissues \cite{dai_quantitative_2022} to advanced optical components such as holographic polarization-controlled metasurfaces \cite{Song_2020}. In previous studies, vectorial ptychography has relied on linear polarization states for both illumination and detection \cite{ferrand_2018}. While this approach has been effective in addressing various challenging situations, it may have limitations when dealing with materials that exhibit strong circular-polarization  properties. Such materials are often encountered in chiral molecular assemblies, which can make reconstruction difficult.

This letter presents an extension to the capabilities of vectorial ptychography, demonstrating how it can be applied to circular-polarizing materials. Initially, we provide theoretical evidence of the underdetermination introduced by linear polarizations when investigating such materials. Subsequently, we propose an enhanced measurement scheme that relies on combinations of elliptical polarizations. We validate this approach through numerical simulations, where the results are analyzed with the help of Pauli matrices algebra. Finally, experimental vectorial ptychography measurements are conducted on cholesteric liquid crystal (CLC) films, which exhibit specific polarization properties.

\section{Theory}

\subsection{Principle of vectorial ptychography}

In vectorial ptychography, the recorded intensity at the $j$-th scanning position is represented as the square modulus of the far field
\begin{equation}
I_{jkl}(\qb) = \left| \mathcal{F} \left( \psi_{jkl}(\rb) \right) \right|^2
\label{Eq:PtychoIntensityGen}
\end{equation}
where where $\psi_{jkl}$ denotes the scalar exit field after analysis for the $k$-th polarized probe and $l$-th polarization analysis \cite{ferrand_2015}. The operator $\mathcal{F}$ represents a propagation operator, while $\rb$ and $\qb$ are the direct and reciprocal space coordinates, respectively. The exit field can be expressed as
\begin{equation}
\psi_{jkl}(\rb) = \hb_l ^\mathrm{t} \Jb(\rb- \rb_j) \pb_k (\rb)
\label{Eq:ExitField}
\end{equation}
where $\hb_l$ is the polarization analysis operator, "t" denotes the transpose operator, $\Jb(\rb- \rb_j)$ represents the Jones matrix map of the laterally shifted investigated object, and $\pb_k (\rb)$ corresponds to the vectorial field distribution of the polarized illumination probe.

The success of the iterative ptychography reconstruction depends on the changes observed in the recorded intensity patterns $I_{jkl}(\qb)$ during spatial ($j$) and polarization ($k, l$) scanning \cite{ferrand_2018}. Previous works have mainly focused on linear polarizations for both the illumination (at angle $\alpha_k$) and analysis (at angle $\theta_l$), given by
\begin{equation}
\pb^\mathrm{lin}_k \propto \begin{bmatrix} \cos \alpha_k \\ \sin \alpha_k \end{bmatrix}
\quad \textrm{and} \quad
\hb^\mathrm{lin}_l \propto \begin{bmatrix} \cos \theta_l \\ \sin \theta_l \end{bmatrix}.
\label{Eq:lin}
\end{equation}
While this measurement scheme has been employed to investigate a wide range of optical properties \cite{baroni_prapp_2020}, the specific case of circular-polarizing materials has not been  studied. 

\subsection{Circular-polarizing materials}

Circular-polarizing materials possess the property of transforming incident light into circularly polarized light, in transmission and/or in reflection. Additionally, if a circular polarization of a particular handedness (left or right) can be transmitted with the same handedness, the circular polarizer is referred to as homogeneous \cite{chipman_polarized_2018}. In this study, it is important to note that we will adopt a convention for which the term "left" refers to a  polarized field that circulates in a counterclockwise direction when observed from the detector's viewpoint. 

\subsection{Linear polarization scheme}

For the sake of simplicity, let us consider a material that behaves as a homogeneous left-circular polarizer in transmission, along with transmittance properties $T(\rb)$ that are independent of polarization. This material can be described by the following Jones matrix,
\begin{equation}
\Jb(\rb) = \frac{1}{2} T(\rb)\begin{bmatrix}1 & -i \\ +i & 1 \end{bmatrix}.
\label{eq:JLCP}
\end{equation}
Under a linear polarization scheme (Eq.~\ref{Eq:lin}), it can be demonstrated that the exit field, as given by Eq.~\ref{Eq:ExitField}, becomes
\begin{equation}
\psi^\mathrm{lin}_{jkl}(\rb) \propto T(\rb- \rb_j) e^{i(\theta_l-\alpha_k)}.
\label{eq:ExitLin}
\end{equation}
It is important to note that the polarization properties contribute to the exit field solely as a spatially homogeneous phase factor in Eq.~\ref{eq:ExitLin}. Consequently, the recorded intensity remains unchanged according to Eq.~\ref{Eq:PtychoIntensityGen}, regardless of the combination of polarizations employed. In a more general context, it can be shown that different types of circular-polarizing materials, whether left or right, homogeneous or inhomogeneous, would yield the same set of diffracted intensities $I_{jkl}(\qb)$ and, therefore, would be indistinguishable using this measurement scheme, compromising the unicity of the ptychographic reconstruction.

\subsection{Improved polarization scheme}

The effective resolution of this ambiguity can be achieved by utilizing a wider range of polarizations.  By substituting the operators $\pb_k$ and $\hb_l$ with the following expressions
\begin{equation}
\pb^\mathrm{ell}_k \propto \begin{bmatrix} \cos \alpha_k \\ - i \sin \alpha_k \end{bmatrix}
\quad \textrm{and} \quad
\hb^\mathrm{ell}_l \propto \begin{bmatrix}  \cos \theta_l \\ - i \sin \theta_l \end{bmatrix},
\label{Eq:ell}
\end{equation}
which correspond to general elliptical polarizations with an azimuth of 0° or 90° and ellipticities defined by $\tan \alpha_k$ and $\tan \theta_l$, respectively, the resulting exit field described by Eq.~\ref{Eq:ExitField}, in the example considered earlier, becomes
\begin{equation}
\psi^\mathrm{ell}_{jkl}(\rb) \propto T(\rb- \rb_j) (\cos \alpha_k - \sin \alpha_k)(\cos \theta_l + \sin \theta_l).
\label{Eq:ExitEll}
\end{equation}
As it will be illustrated later, the concept can be easily grasped by considering the Poincaré sphere representation for polarizations. It involves investigating multiple points on a sphere's meridian for both illumination and analysis, involving spherical, elliptical, and linear polarizations. This situation allows us to overcome the barrier imposed by linear polarizations positioned on the sphere's equator. This modification introduces a clear signature of the circular polarization properties, resulting in a change in the amplitude of the exit field as given by Eq.~\ref{Eq:ExitEll}, depending on the specific combination of polarizations. Furthermore, it can be demonstrated that the amplitude factors in Eq.~\ref{Eq:ExitEll} take different algebraic forms, facilitating the unequivocal identification of the type of circular polarizer. The relevance of this latter approach has been tested on both numerical and experimental datasets.

\section{Materials and methods}

\subsection{Numerical simulations}

Numerical datasets were generated to simulate an experiment of vectorial ptychography at a wavelength $\lambda = 635$~nm using 50-µm-diameter cropped Gaussian-shaped illumination probes under which an object is raster-scanned with a step size of  7~µm in both directions, resulting in an $11\times11$ grid. Far-field intensity patterns were computed as if they were captured at an infinite distance, within a numerical aperture of 0.3, and recorded on a camera sensor with dimensions of $122\times122$ pixels. To replicate the effects of shot noise on the sensor, a Poisson random number generator was employed. Typically, each frame had a total photon count of $10^6$. The datasets were processed using the vectorial ptychographic iterative algorithm detailed in a previous publication \cite{ferrand_2015}. The algorithm employed random distributions (modulus and phase) for the initial guesses and was run for 30~iterations with known probes.

\subsection{CLC films}

Cholesteric liquid crystals (CLC) films were experimentally investigated at $\lambda=635$~nm. These materials are characterized by a helical structure of pitch $p$ that produces a specific optical response for wavelengths within a bandwidth centered at $\lambda_0=\bar{n}p$, at normal incidence, where $\bar{n}$ is the mean refractive index. The reflected light is circularly polarized. When unpolarized light is incident on a CLC, a maximum of 50\% of light is reflected (matching the helix handedness), and 50\% of the light is transmitted (circularly polarized with the inverse handedness) \cite{mitov_cholesteric_2012}. In the following, this case will be referred to as "Bragg film", in analogy with X-ray diffraction. On the contrary, if the wavelength $\lambda$ is outside the bandwidth, the polarization rule is ineffective and the material will be referred to as "off-Bragg". Wacker-Chemie GmbH provided us with CLCs polysiloxane-based oligomers. Details regarding their chemical and physical properties, as well as relevant references, are given in a prior paper \cite{scarangella_biomimetic_2020}. Their glass transition temperature $T_\textrm{g}$ ranges from 40 to 55°C. The helix is left-handed. Thin films were produced between two plain glass substrates and annealed at 140°C in their viscous state. They can be vitrified by quenching below $T_\textrm{g}$, and their cholesteric structure is preserved at normal temperature in a solid state. The Silicon-Green chemical was used to fabricate the off-Bragg film (half height bandwidth 430--500 nm). It was annealed for 90 min without the top substrate, resulting in a blue-shift in the reflection band \cite{agez_color_2011}. The thickness measures $2.6 \pm 0.1$~µm. The Bragg film (half height bandwidth 550--680 nm) is a 13:87 wt. \% mixture of Silicon-Blue and Silicon-Red compounds. It was annealed for 10 min with both substrates present. The thickness measures $15\pm2$~µm.

\subsection{Vectorial ptychography measurements}

Measurements were carried out at $\lambda = 635$~nm on an optical setup described previously \cite{ferrand_2018}, adapted here by inserting two quarter waveplates properly oriented, one before the object, one after, as shown in Fig.~S1 in Supplement 1. This modification provides a simple solution to upgrade the linear polarization scheme to the improved one proposed in this work. Thus, the mechanical control of the polarization, formerly of the angles of linear polarizations (Eqs.~\ref{Eq:lin}) acts now as a control of the polarization ellipticities (Eqs.~\ref{Eq:ell}). Measurements were carried out with an illumination probe of effective diameter 100~µm.
The far-field was collected through a numerical aperture of 0.4 and recorded by a camera of effective dimensions $320 \times 240$ pixels. Object reconstructions were performed by means of 500 iterations of a conjugate-gradient algorithm described in a previous work \cite{baroni_joint_2019}, allowing the joint estimation of the three probes together with the Jones maps of the object, with a pixel size of about $0.73 \times 0.97$~µm$^2$.

\section{Results and discussion}

\begin{figure}[htbp]
\begin{center}
\includegraphics[width=.85\linewidth]{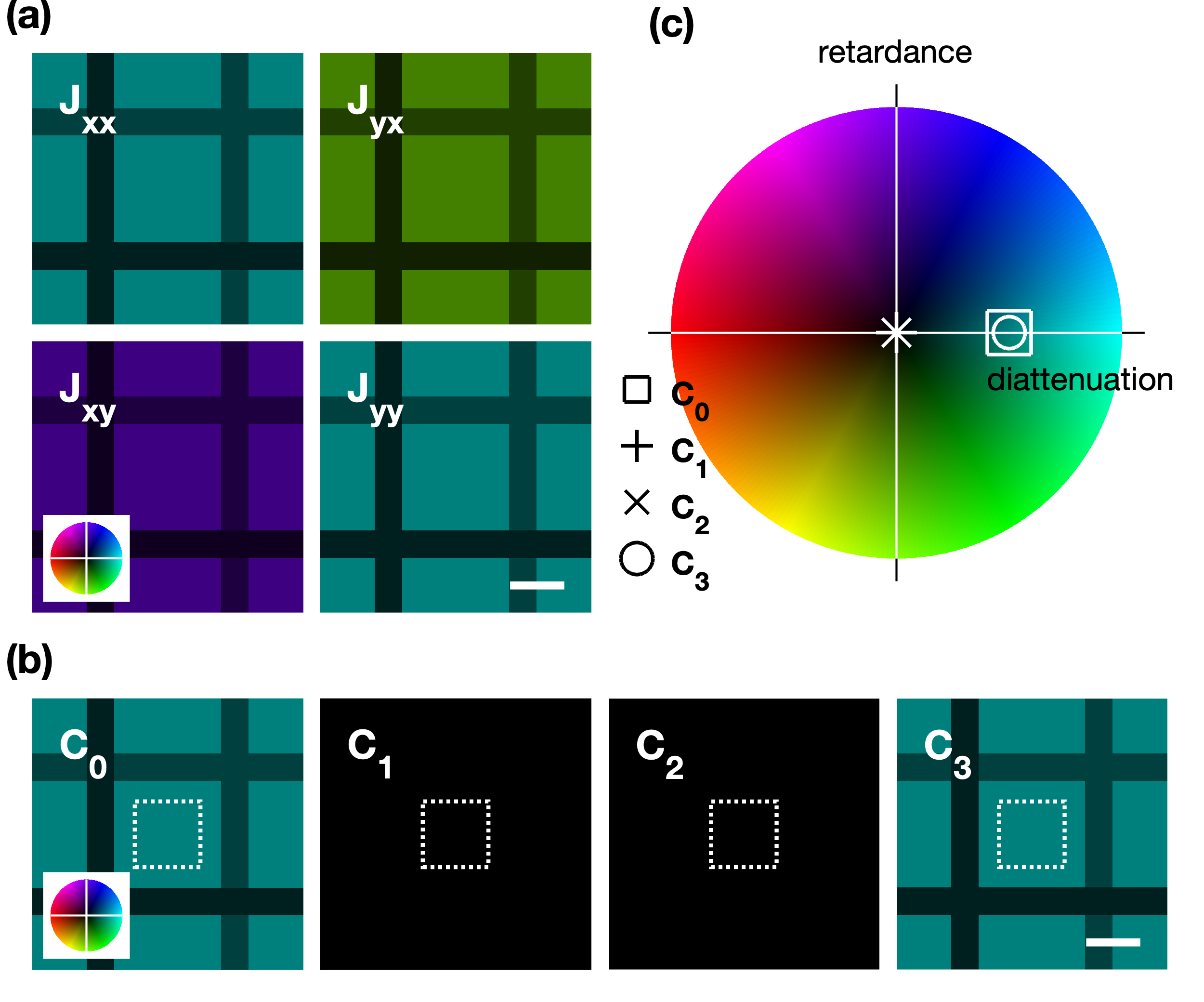}
\end{center}
\caption{Description of the simulated numerical object. (a) Jones maps $\Jb(\rb)$. Inset show the colour encoding for complex values in the complex plane. (b) Pauli coefficients maps. (c) Values of the complex Pauli coefficients represented in the complex plane. Values have been averaged in the square dotted area shown on the maps of panel b. Scale bars are 20~µm.}
\label{fig:NumObj}
\end{figure}

We performed a simulation using an object described by the Jones matrix presented in Eq.~\ref{eq:JLCP}. The transmittance properties $T(\rb)$ of the object corresponded to a uniform region intersected by vertical and horizontal stripes with lower transmittance. The resulting Jones maps are depicted in Fig.~\ref{fig:NumObj}a. To better identify the presence of circular polarization properties while preserving all information, we expanded them as a sum
\begin{equation}
\Jb(\rb)= \sum_{n=0}^3 C_n(\rb) \ \pmb{\sigma}_n.
\end{equation}
Here, $\pmb{\sigma}_0$ denotes the identity matrix, and $\pmb{\sigma}_1$, $\pmb{\sigma}_2$, and $\pmb{\sigma}_3$ represent the Pauli matrices \cite{chipman_polarized_2018}
\begin{equation}
\pmb{\sigma}_1 = 
\begin{bmatrix}
1 & 0 \\
0 & -1
\end{bmatrix} ; \ 
\pmb{\sigma}_2 = 
\begin{bmatrix}
0 & 1 \\
1 & 0
\end{bmatrix} ; \
\pmb{\sigma}_3 = 
\begin{bmatrix}
0 & -i \\
i & 0
\end{bmatrix}.
\end{equation}
The corresponding maps of the four complex Pauli coefficients $C_n(\rb)$ are displayed in Fig.~\ref{fig:NumObj}b. Their average values, obtained near the center of the image, are illustrated in the complex plane (Fig.~\ref{fig:NumObj}c). These values highlight the homogeneous left-circular polarizer properties of the simulated object, where $C_0$ and $C_3$ are equal, positive, and real, while $C_1=C_2=0$.

\begin{figure}[!h]
\begin{center}
\includegraphics[width=.9\linewidth]{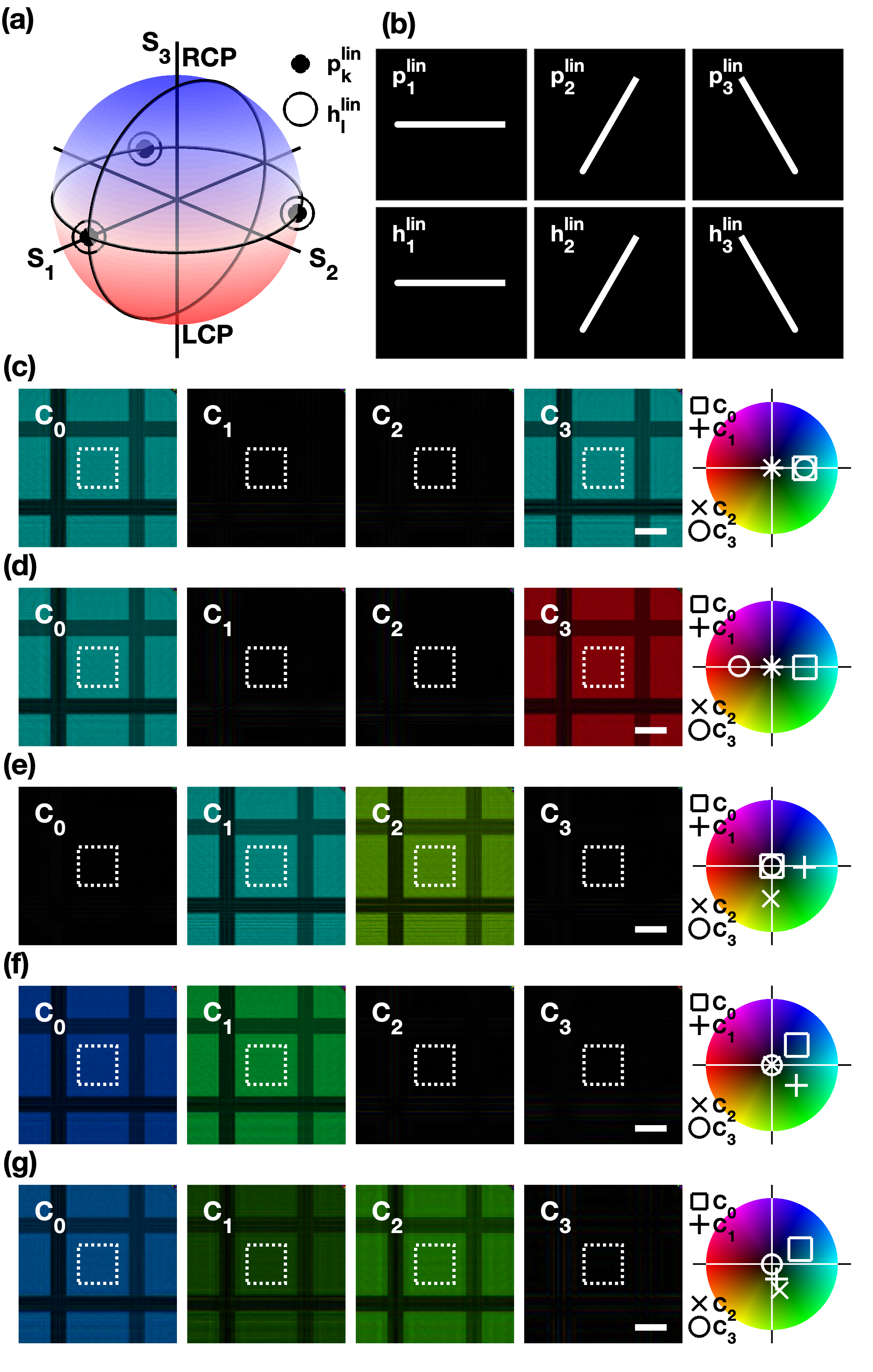}
\end{center}
\caption{Simulations of measurements with the linear polarization scheme. (a) Combinations of polarizations for illuminations ($\pb^\mathrm{lin}_k$, solid circle) and analyses ($\hb^\mathrm{lin}_l$, empty circle) shown on the Poincaré sphere. $S_1$, $S_2$ and $S_3$ are the standard Stokes coefficients \cite{chipman_polarized_2018}. RCP and LCP stand for right and left circular polarization, respectively. (b) Illustration of these polarizations. The colour indicates the ellipticity, given by the latitude on the Poincaré sphere. (c-g) Pauli coefficient maps calculated from retrieved Jones maps, for several reconstruction, run independently, represented with the same convention as in Figs.~\ref{fig:NumObj}b and c. Scale bars are 20~µm.}
\label{fig:LinearNum} 
\end{figure}

First, we conducted a simulation to replicate a measurement using the linear-polarization scheme. The specific angles employed were $\alpha_1=0$°, $\alpha_2=60$°, $\alpha_3=120$°, along with $\theta_1=0$°, $\theta_2=60$°, $\theta_3=120$° \cite{Song_2020} in Eqs.~\ref{Eq:lin}, as visually depicted in Fig.~\ref{fig:LinearNum}a,b. Multiple independent algorithm runs were executed, yielding diverse solutions. Figs~\ref{fig:LinearNum}c-e provide an overview by displaying maps of the Pauli coefficients for each solution. Alongside the correct solution (Fig.~\ref{fig:LinearNum}c), which indicated a homogeneous left-circular polarizer, several alternative solutions emerged. These alternative solutions comprised of either homogeneous right-circular polarizers (Fig.~\ref{fig:LinearNum}d) or inhomogeneous ones (Fig.\ref{fig:LinearNum}e), thereby confirming the theoretical ambiguity highlighted. Remarkably, the discovered solutions extended beyond circular-polarizing characteristics, exhibiting different properties. For example, Fig.~\ref{fig:LinearNum}f corresponded to a horizontal quarter waveplate, while the last case, Figs.~\ref{fig:LinearNum}g, represented more intricate combinations.

\begin{figure}[!t]
\includegraphics[width=\linewidth]{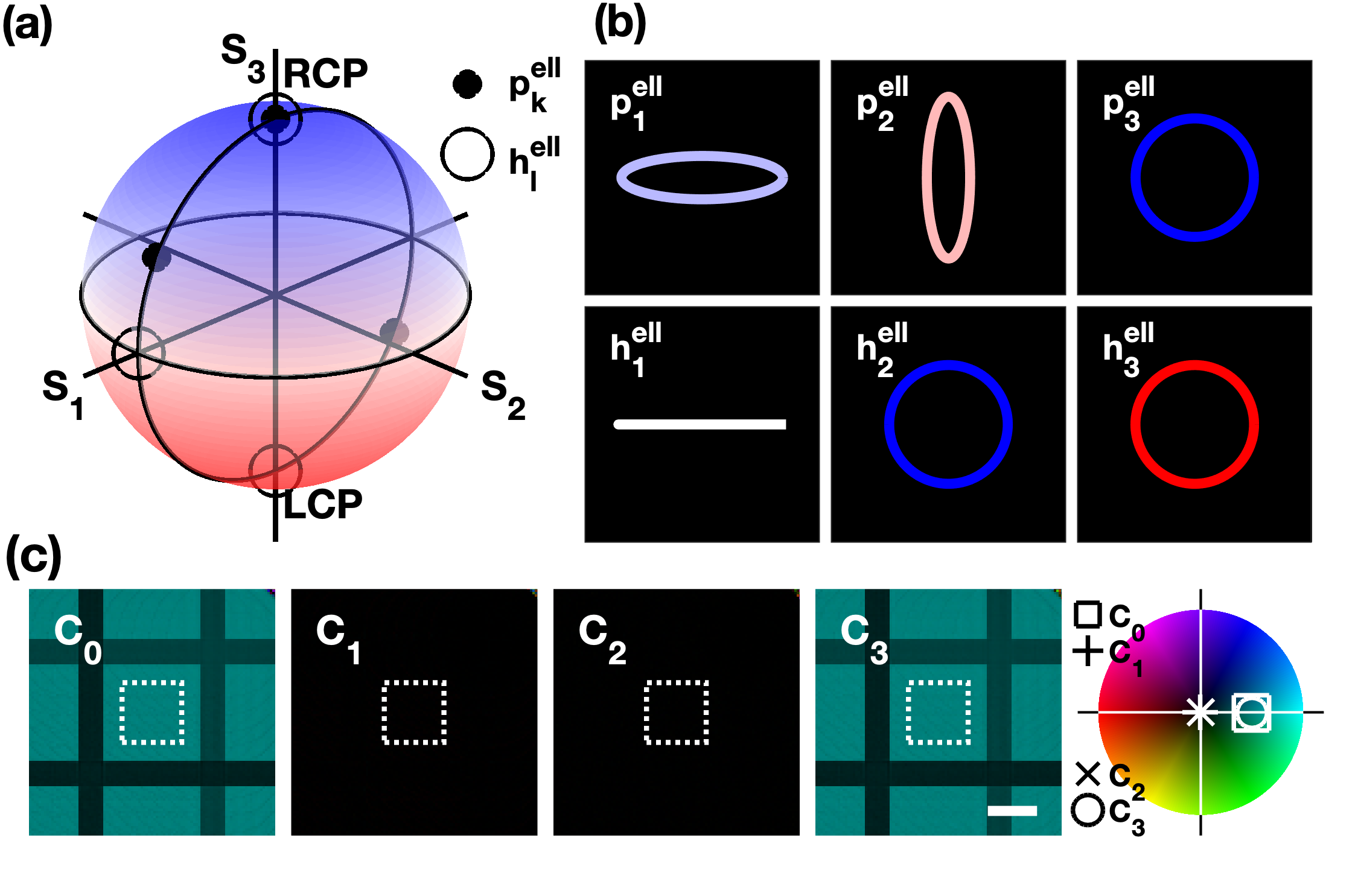}
\caption{Simulations of measurements with the improved polarization scheme. (a) Combinations of polarizations for illuminations ($\pb^\mathrm{ell}_k$, solid circle) and analyses ($\hb^\mathrm{ell}_l$, empty circle) shown on the Poincaré sphere. (b) Corresponding polarizations. (c) Pauli coefficient maps calculated from retrieved Jones maps, as obtained systematically for any reconstruction run independently, represented with the same convention as in Figs.~\ref{fig:NumObj}b and c. Scale bars are 20~µm.}
\label{fig:EllNum}
\end{figure}

Subsequently, the simulation was executed using an improved polarization scheme. The chosen angles were $\alpha_1=15$°, $\alpha_2=105$°, $\alpha_3=45$°, in conjunction with $\theta_1=0$°, $\theta_2=-45$°, $\theta_3=45$° within Eqs.~\ref{Eq:ell}, visually represented in Fig.~\ref{fig:EllNum}a. These angles, chosen empirically, represented a combination of linear, elliptical, and circular polarization states, allowing to probe efficiently a large variety of optical properties. Once again, multiple runs of the algorithm were performed, resulting in consistent convergence towards the correct solution, now. The obtained solution is illustrated in Fig.~\ref{fig:EllNum}b, further validating the effectiveness of this scheme.

\begin{figure}[!h]
\includegraphics[width=\linewidth]{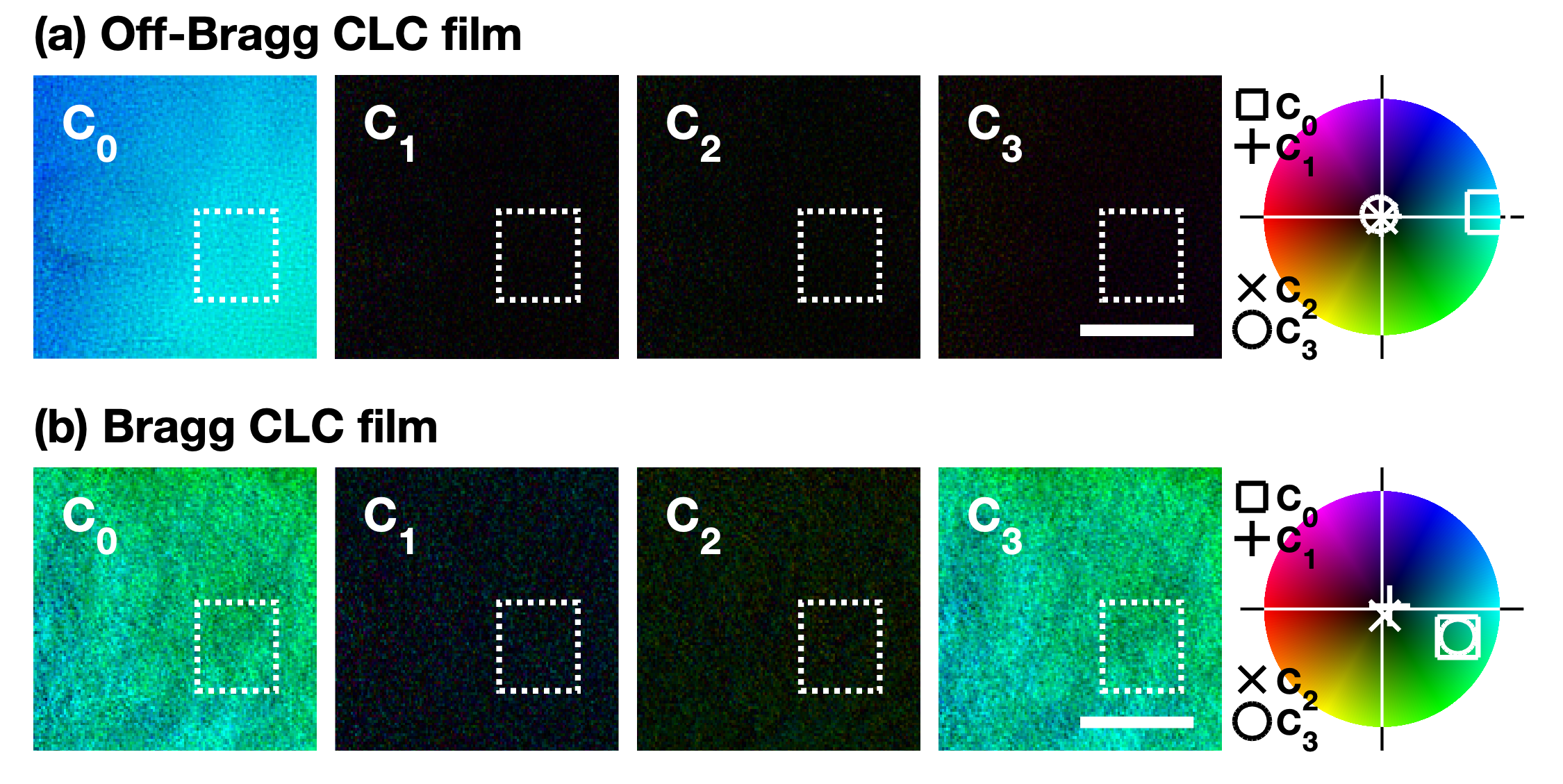}
\caption{Investigation of two CLC films. Pauli coefficient maps. (a) Off-Bragg film. (b) Bragg film. Scale bars are 40~µm.}
\label{fig:CholReco}
\end{figure}

Finally, Bragg and off-Bragg CLC films were investigated using vectorial ptychography with the improved polarization scheme. Fig.~\ref{fig:CholReco}a illustrates the Jones and Pauli coefficients maps obtained from the reconstruction algorithm. For the off-Bragg film (Fig.~\ref{fig:CholReco}a), the Pauli coefficients maps show that only the $C_0(\rb)$ map has non-zero values. This confirms that the film has an optical response that is insensitive to polarization. On the contrary, the Bragg film (Fig.~\ref{fig:CholReco}b) displays different maps of Pauli coefficients, with $C_0(\rb) = C_3(\rb) \neq 0$, while $C_1(\rb)=C_2(\rb)=0$. This confirms that the film functions as a uniform left-circular polarizer. Furthermore, variations in the Pauli coefficient allow to report a texturing of the films (polygonal texture)  that has been discussed previously \cite{agez_color_2011}. Addressing this aspect is beyond the scope of the present study and will be the subject of a future article. It is worth emphasizing the clear advantage of spatial resolution and intrinsic phase-imaging capabilities offered by vectorial ptychography \cite{baroni_prapp_2020}. In contrast, other interferometric methods typically provide spatially averaged measurements \cite{sanchez-castillo_circular_2014}.

\section{Conclusion}

In conclusion, this letter presents a novel approach to vectorial ptychography, expanding its capabilities to circular-polarizing materials such as CLCs. By a measurement scheme based on elliptical polarizations, the technique overcomes the limitations of linear polarizations and enables the accurate characterization of circular polarizers at a micrometer-scale resolution. This progress in quantitative imaging has implications for various fields, including the study of chiral molecular assemblies and other materials or advanced components with strong circular-polarization  properties.

\begin{backmatter}
\bmsection{Funding} We acknowledge funding from the European Research Council (ERC) under the European
Union’s Horizon 2020 research and innovation program (Grant agreement no. 724881).

\bmsection{Acknowledgments} We thank Dr. E. Hanelt from Wacker-Chemie GmbH (Munich, Germany) for providing us with cholesteric oligomers.

\bmsection{Disclosures} The authors declare no conflicts of interest.

\bmsection{Data Availability Statement} The datasets  are available upon reasonable request from the corresponding author.

\bmsection{Supplemental document}
See Supplement 1 for supporting content. 

\end{backmatter}


\bibliography{myrefs}

\begin{thebibliography}{10}
\newcommand{\enquote}[1]{``#1''}

\bibitem{ferrand_2015}
P.~Ferrand, M.~Allain, and V.~Chamard, \enquote{Ptychography in anisotropic
  media,} {\protect\JournalTitle{Optics Letters}} \textbf{40}, 5144 (2015).

\bibitem{ferrand_2018}
P.~Ferrand, A.~Baroni, M.~Allain, and V.~Chamard, \enquote{Quantitative imaging
  of anisotropic material properties with vectorial ptychography,}
  {\protect\JournalTitle{Optics Letters}} \textbf{43}, 763 (2018).

\bibitem{wang_optical_2023}
T.~Wang, S.~Jiang, P.~Song, R.~Wang, L.~Yang, T.~Zhang, and G.~Zheng,
  \enquote{Optical ptychography for biomedical imaging: recent progress and
  future directions [{Invited}],} {\protect\JournalTitle{Biomedical Optics
  Express}} \textbf{14}, 489--532 (2023).

\bibitem{baroni_prapp_2020}
A.~Baroni, V.~Chamard, and P.~Ferrand, \enquote{Extending {Quantitative}
  {Phase} {Imaging} to {Polarization}-{Sensitive} {Materials},}
  {\protect\JournalTitle{Physical Review Applied}} \textbf{13}, 054028 (2020).

\bibitem{duboisset_2022}
J.~Duboisset, P.~Ferrand, A.~Baroni, T.~A. Grünewald, H.~Dicko, O.~Grauby,
  J.~Vidal-Dupiol, D.~Saulnier, L.~M. Gilles, M.~Rosenthal, M.~Burghammer,
  J.~Nouet, C.~Chevallard, A.~Baronnet, and V.~Chamard,
  \enquote{Amorphous-to-crystal transition in the layer-by-layer growth of
  bivalve shell prisms,} {\protect\JournalTitle{Acta Biomaterialia}}
  \textbf{142}, 194--207 (2022).

\bibitem{dai_quantitative_2022}
X.~Dai, S.~Xu, X.~Yang, K.~C. Zhou, C.~Glass, P.~C. Konda, and R.~Horstmeyer,
  \enquote{Quantitative {Jones} matrix imaging using vectorial {Fourier}
  ptychography,} {\protect\JournalTitle{Biomedical Optics Express}}
  \textbf{13}, 1457--1470 (2022).

\bibitem{Song_2020}
Q.~Song, A.~Baroni, R.~Sawant, P.~Ni, V.~Brandli, S.~Chenot, S.~Vézian,
  B.~Damilano, P.~de~Mierry, S.~Khadir, P.~Ferrand, and P.~Genevet,
  \enquote{Ptychography retrieval of fully polarized holograms from
  geometric-phase metasurfaces,} {\protect\JournalTitle{Nature Communications}}
  \textbf{11}, 2651 (2020).

\bibitem{chipman_polarized_2018}
R.~A. Chipman, W.~S.~T. Lam, G.~Young, W.~S.~T. Lam, and G.~Young,
  \emph{Polarized {Light} and {Optical} {Systems}} (CRC Press, 2018).

\bibitem{mitov_cholesteric_2012}
M.~Mitov, \enquote{Cholesteric {Liquid} {Crystals} with a {Broad} {Light}
  {Reflection} {Band},} {\protect\JournalTitle{Adv. Mater.}} \textbf{24},
  6260--6276 (2012).

\bibitem{scarangella_biomimetic_2020}
A.~Scarangella, V.~Soldan, and M.~Mitov, \enquote{Biomimetic design of
  iridescent insect cuticles with tailored, self-organized cholesteric
  patterns,} {\protect\JournalTitle{Nature Commun.}} \textbf{11}, 4108 (2020).

\bibitem{agez_color_2011}
G.~Agez, R.~Bitar, and M.~Mitov, \enquote{Color selectivity lent to a
  cholesteric liquid crystal by monitoring interface-induced deformations,}
  {\protect\JournalTitle{Soft Matter}} \textbf{7}, 2841--2847 (2011).

\bibitem{baroni_joint_2019}
A.~Baroni, M.~Allain, P.~Li, V.~Chamard, and P.~Ferrand, \enquote{Joint
  estimation of object and probes in vectorial ptychography,}
  {\protect\JournalTitle{Opt. Exp.}} \textbf{27}, 8143 (2019).

\bibitem{sanchez-castillo_circular_2014}
A.~Sanchez-Castillo, S.~Eslami, F.~Giesselmann, and P.~Fischer,
  \enquote{Circular polarization interferometry: circularly polarized modes of
  cholesteric liquid crystals,} {\protect\JournalTitle{Opt. Exp.}} \textbf{22},
  31227--31236 (2014).

\end{thebibliography}

\bibliographyfullrefs{myrefs}

\end{document}